\documentclass[aps,pra,twocolumn,showpacs,floatfix]{revtex4}
\usepackage{epsfig}
\usepackage{graphicx}
\usepackage{dcolumn}
\usepackage{amsthm,amsmath}

\begin{document}

\title{Determination of magic wavelengths for the $7s ~ {^2}S_{1/2}-7p ~ {^2}P_{3/2,1/2}$ transitions in Fr atom}

\author{Sukhjit Singh$^a$\footnote{Email: sukhjitphy.rsh@gndu.ac.in}, B. K. Sahoo$^{b}$\footnote{Email: bijaya@prl.res.in}, 
and Bindiya Arora$^a$\footnote{Email: bindiya.phy@gndu.ac.in} }
\affiliation{$^a$Department of Physics, Guru Nanak Dev University, Amritsar, Punjab-143005, India}
\affiliation{$^b$Atomic, Molecular and Optical Physics Division, Physical Research Laboratory, Navrangpura, Ahmedabad-380009, India}

\date{Received date; Accepted date}

\begin{abstract}
Magic wavelengths ($\lambda_{\rm{magic}}$) for the $7S_{1/2}-7P_{1/2,3/2}$ transitions (D-lines) in Fr were reported 
by Dammalapati \textit{et al.} in [Phys. Rev. A 93, 043407 (2016)]. These $\lambda_{\rm{magic}}$ were determined by plotting 
dynamic polarizabilities ($\alpha$) of the involved states with the above transitions against a desired range of wavelength. 
Electric dipole (E1) matrix elements listed in [J. Phys. Chem. Ref. Data 36, 497 (2007)], from the measured lifetimes of 
the $7P_{1/2,3/2}$ states and from the calculations considering core-polarization effects in the relativistic 
Hartree-Fock (HFR) method, were used to determine $\alpha$. However, contributions from core correlation effects and 
from the E1 matrix elements of the $7P-7S$, $7P-8S$ and $7P-6D$ transitions to $\alpha$ of the $7P$ states were ignored. 
In this work, we demonstrate importance of these contributions and improve accuracies of $\alpha$ further by replacing 
the E1 matrix elements taken from the HFR method by the values obtained employing relativistic coupled-cluster theory. 
Our static $\alpha$ are found to be in excellent agreement with the other  available theoretical results; whereas 
substituting the E1 matrix elements used by Dammalapati \textit{et al.} give very small $\alpha$ values for the $7P$ states. Owing
to this, we find disagreement in $\lambda_{\rm{magic}}$ reported by Dammalapati \textit{et al.} for linearly polarized light; 
especially at wavelengths close to the D-lines and in the infrared region. As a consequence, a $\lambda_{\rm{magic}}$ 
reported at 797.75 nm which was seen supporting a blue detuned trap in their work is now estimated at 771.03 nm and is supporting a red 
detuned trap. Also, none of our results match with the earlier results for circularly polarized light. Moreover, our 
static values of $\alpha$ will be very useful for guiding experiments to carry out their measurements. 
\end{abstract}

\maketitle

\section{Introduction}

Being the heaviest alkali atom, Fr atom is considered for measuring electric dipole moment (EDM) due to parity 
and time reversal symmetries \cite{sakemi,inou,Mukherjee10}, parity nonconservation (PNC) effect in the 
$7s \ ^2S_{1/2} \rightarrow 8s \ ^2S_{1/2}$ transition due to neutral weak interaction \cite{stancari,bijaya1} and PNC
effect among the hyperfine transitions in the ground state \cite{gomez1} and in the $7s \ ^2S_{1/2} \rightarrow 6d \ ^2D_{5/2}$ 
transition due to the nuclear anapole moment \cite{bijaya2}. Recent theoretical studies on hyperfine structures in $^{210}$Fr and 
$^{212}$Fr demonstrate inconsistencies between the theoretically evaluated and measured hyperfine structure constants of few 
excited states \cite{ref1}. The hyperfine structure constants and lifetimes of the $6d \ ^2D_{3,5/2}$ states of Fr, which are important 
for PNC studies \cite{ref1,ref2}, have not been measured yet. Also, suggestion to measure hyperfine splitting in the suitable transitions
to observe nuclear octupole moment of its $^{211}$Fr isotope have been made \cite{bijaya3}. To carry out high precision measurements 
for all the above mentioned vital studies, it is indispensable to conduct experiments on Fr atoms in an environment where they are least
affected by external perturbations. In such scenario, performing experiments by cooling and trapping Fr atoms using lasers can be 
advantageous. To estimate the induced Stark shifts in the energy levels due to the applied lasers, knowledge about precise values of 
polarizabilities is necessary. There are no experimental results on polarizabilities in Fr available yet, while only 
a few theoretical results are reported.

Techniques to produce Fr atoms and trapping them in a magneto-optic trap (MOT) have already been demonstrated \cite{Simsarian2, atutov}. Similar to other alkali atoms D-lines in Fr atom are used for laser cooling and trapping experiments, which leads to the easy accessibility of this atom for its application in probing new physics of fundamental particles~\cite{Dammalapati,phillip,Ghosh}. Therefore, it is certainly attainable to develop cooling and trapping techniques for the Fr 
atoms in near future. It is worth mentioning here that recently there have been proposals suggesting to adopt these methodologies to 
measure PNC and EDM in the Fr atom \cite{bijaya2,inou}. However, when lasers are applied to the atoms, the Stark shifts experienced
by the energy levels cause large systematics to carry out high precision measurements of any spectroscopic properties. One of the most 
innovative ways to circumvent this problem is to trap the atoms at the magic wavelengths ($\rm\lambda_{magic}$) at which differential 
Stark shift of a transition is effectively nullified. The concept of $\rm\lambda_{magic}$ was first introduced by Katori {\it et al.} 
for its application in the optical atomic clocks \cite{Katori11}. In fact, $\rm\lambda_{magic}$ of the $6s ~{^2}S_{1/2}$-$6p ~{^2}P_{3/2}$ 
transition in Cs has been measured by Mckeever \textit{et al.} at 935.6 nm \cite{mck} using linearly polarized light. In our previous 
works, we have also theoretically determined $\rm\lambda_{magic}$ of D-lines of the lighter alkali atoms for both linearly and 
circularly polarized light \cite{arora1,Arora86,bijaya4,sukhjit1,sukhjitnew}. With the same objective, Dammalapati \textit{et al.} \cite{Dammalapati} 
have recently identified $\rm\lambda_{magic}$ for the $7S_{1/2}-7P_{1/2,3/2}$ and $7S_{1/2}-8S_{1/2}$ transitions in Fr 
considering both linearly and circularly polarized light. On this rationale, they have used transition rates compiled by Sansonetti in 
Ref. \cite{Sansonetti} to calculate the required dynamic dipole polarizabilities. Few of these transition probabilities quoted by 
Sansonetti were extracted from the measurements of the lifetimes of the $7P_{1/2,3/2}$ states of Fr, while the remaining data were 
taken from the calculations, based on the relativistic Hartree-Fock (HFR) method accounting only the core-polarization effects,  
carried out by Biemont {\it et al.} \cite{biemont}. However, these calculations of polarizabilities completely ignore contributions 
coming from the correlations due to the core electrons (known as core correlation contribution), which are about 6\% in the evaluation 
of the static values of polarizabilities as has been demonstrated later, and correlations among the core and valence electrons 
(referred in the literature as core-valence correlation contribution), and from the high-lying transitions (tail contribution) involving
states above $n=20$, for the principal quantum number $n$. Most importantly, Dammalapati \textit{et al.} have not considered the contributions from the 
$7P-7S$, $7P-8S$ and $7P-6D$ transitions in their calculations of the dynamic polarizabilities of the $7P_{1/2,3/2}$ states. As 
shown in this work subsequently, contributions from these states are more than 80\% to the static polarizability values of the 
$7P_{1/2,3/2}$ states. Therefore, it is imperative to determine $\rm\lambda_{magic}$ of the important D-lines of the Fr atom more 
precisely by determining polarizabilities of the atomic states more accurately. 

We pursue this work intending to improve evaluation of $\rm\lambda_{magic}$ over the previously reported values for both linearly and circularly 
polarized light by including the core, core-valence and tail contributions and also accounting contributions from the $7P-7S$, $7P-8S$ 
and $7P-6D$ transitions. In addition to this, we use more accurate values of the electric dipole (E1) matrix elements for the higher 
excited states from a relativistic coupled-cluster (RCC) theory as compared to the values used from a lower order many-body method in 
\cite{Dammalapati}. In order to validate our results, we have also estimated static dipole polarizability values of the ground and 
$7P_{1/2,3/2}$ states and compare them against the other available high precision calculations. In order to demonstrate importance of 
inclusion of the appended contributions in the evaluation of the polarizabilities and also to find out possible reason for the  
discrepancies in the $\rm\lambda_{magic}$ from both the works, we present contributions to the static polarizabilities from various 
transitions, core correlations and core-valence correlations explicitly. We also estimate the static polarizability values of the above
states obtained exclusively using the E1 matrix elements considered by Dammalapati \textit{et al.} and compare them with the other theoretical 
results.

\begin{table*}
\caption{\label{pol1}Contributions from reduced E1 matrix elements (given as $d$), core correlation and core-valence correlation to the static polarizabilities of the 
$7S_{1/2}$, $7P_{1/2}$ and $7P_{3/2}$ states of Fr atom. The final results are compared with the previously estimated results. The
results marked `*' are calculated using the  reduced E1 matrix elements (taken from Ref.~\cite{Sansonetti}) which are considered in
Ref.~\cite{Dammalapati} to determine $\rm\lambda_{magic}$. All the values are given in atomic units (a.u.).}
%\begin{ruledtabular}
\begin{center}
\begin{tabular}{|ccc|ccc|cccc|}
\hline 
\hline
\multicolumn{3}{|c|}{$7S_{1/2}$ state}  & \multicolumn{3}{c|}{$7P_{1/2}$ state}  &  \multicolumn{4}{c|}{$7P_{3/2}$ state}\\
& & & & & & & & & \\ 
 Transition         & $d$ &  $\alpha_n^{(0)}$ & Transition &  $d$ & $\alpha_n^{(0)}$ &  
 Transition  &  $d$ &   $\alpha_n^{(0)}$ & $\alpha_n^{(2)}$ \\
 
 \hline
 & & & & & & & & & \\ 
 $7S_{1/2}-7P_{1/2}$  &  4.277     & 109.36    & $7P_{1/2}-7S_{1/2}$  &   4.277  & -109.36  & $7P_{3/2}-7S_{1/2}$       &   5.898    & -91.39  &   91.39           \\
 $7S_{1/2}-8P_{1/2}$  &  0.33      & 0.35      & $7P_{1/2}-8S_{1/2}$  &   4.27   &  177.79  & $7P_{3/2}-8S_{1/2}$       &   7.52     &  355.67  &  -355.67          \\
 $7S_{1/2}-9P_{1/2}$  &  0.11      & 0.03      & $7P_{1/2}-9S_{1/2}$  &   1.02   &    5.67  & $7P_{3/2}-9S_{1/2}$       &   1.39     &  6.02    &  -6.02         \\
 $7S_{1/2}-10P_{1/2}$ &  0.06      & 0.01      & $7P_{1/2}-10S_{1/2}$ &   0.54   &    1.33  & $7P_{3/2}-10S_{1/2}$      &   0.71     &  1.28    &  -1.28          \\ 
 $7S_{1/2}-11P_{1/2}$ &  0.04      & $\sim$0   & $7P_{1/2}-11S_{1/2}$ &   0.35   &    0.51  & $7P_{3/2}-11S_{1/2}$      &   0.45     & 0.47    &  -0.47   \\ 
 $7S_{1/2}-7P_{3/2}$  &  5.898     & 182.77    & $7P_{1/2}-6D_{3/2}$  &   7.45   & 1017.03  & $7P_{3/2}-6D_{3/2}$       &   3.44     &  187.72  &  150.18                 \\
 $7S_{1/2}-8P_{3/2}$  &  0.95      & 2.79      & $7P_{1/2}-7D_{3/2}$  &   3.27   &   65.15  & $7P_{3/2}-7D_{3/2}$       &   2.07     &  15.19   &   12.15            \\
 $7S_{1/2}-9P_{3/2}$  &  0.44      & 0.52      & $7P_{1/2}-8D_{3/2}$  &   1.79   &   15.26  & $7P_{3/2}-8D_{3/2}$       &   1.00     &  2.67    &   2.14         \\ 
 $7S_{1/2}-10P_{3/2}$ &  0.28      & 0.19      & $7P_{1/2}-9D_{3/2}$  &   1.17   &    5.86  & $7P_{3/2}-9D_{3/2}$       &   0.62     &   0.91    &   0.73       \\ 
 $7S_{1/2}-11P_{3/2}$ &  0.18      & 0.08      & $7P_{1/2}-10D_{3/2}$ &   0.84   &    2.86  & $7P_{3/2}-10D_{3/2}$      &   0.44     &   0.43    &   0.35  \\
                      &            &           &                      &          &          & $7P_{3/2}-6D_{5/2}$       &   10.53    & 1618.72 &  -323.74\\ 
                      &            &           &                      &          &          & $7P_{3/2}-7D_{5/2}$       &    5.91    & 122.74  &   -24.55\\ 
                      &            &           &                      &          &          & $7P_{3/2}-8D_{5/2}$       &    2.91    & 22.57   &    -4.51\\ 
                      &            &           &                      &          &          & $7P_{3/2}-9D_{5/2}$       &    1.83    & 7.95    &    -1.59\\ 
                      &            &           &                      &          &          & $7P_{3/2}-10D_{5/2}$      &    1.27    & 3.60    &    -0.72\\  
 & & & & & & & & & \\
  Main($\rm \alpha_{n,v}$)  &               &  296.10 &  Main($\rm \alpha_{n,v}$)                   &               & 1182.10  &  Main($\rm \alpha_{n,v}$)                         &                & 2254.56   &  -461.62               \\
  Tail($\rm \alpha_{n,v}$)  &               & 1.26   &   Tail($\rm \alpha_{n,v}$)                    &               & 22.89     &   Tail($\rm \alpha_{n,v}$)                         &                &  29.15     &  -5.24            \\
  $\rm \alpha_{n,cv}$    &               & -0.95  &  $\rm \alpha_{n,cv}$                    &               & $\sim$0    &  $\rm \alpha_{n,cv}$                         &                & $\sim$0     &   $\sim 0$          \\
$\rm \alpha_{n,c}$    &               & 20.4   &    $\rm \alpha_{n,c}$                  &               & 20.4    &   $\rm \alpha_{n,c}$                         &                & 20.4     &                  \\
 Total   &               & 316.81  &      Total                 &               & 1225.39   &       Total                     &                & 2304.10    &  -466.86                  \\
& & & & & & & & & \\
 Others    &               & 317.8(2.4) \cite{Derevianko10} & Others       &               &1106 \cite{Wijngaarden}    & Others &       & 2102.6 \cite{Wijngaarden}       & -402.76 \cite{Wijngaarden}    \\
     &                      & 315.2 \cite{Lim10}  &                      &               &  57.23*  &                        &                & 98.57*    &    63.23*             \\
           &               & 289.8* &                    &               &   &  &       &        &     \\

\hline
\end{tabular}
\end{center}
%\end{ruledtabular}
\end{table*} 

\section{Theory}

The Stark shift in the energy level of an atom in state $|\gamma _n J_n M_{J_n}\rangle$ placed in an uniform oscillating electric field 
$\mbox{\boldmath${\cal E}$}(t)= \frac{1}{2}{\cal E} \hat{\mbox{\boldmath$\varepsilon$}} e^{-\iota\omega t}+c.c.$, with ${\cal E}$ 
being the amplitude, $\hat{\mbox{\boldmath$\varepsilon$}}$ is the polarization vector of the electric field and $c.c.$ referring to the complex conjugate of the former term, 
oscillating at frequency $\omega$ is given by \cite{bonin,manakov,beloy}
\begin{eqnarray}
\label{eq1}
\Delta E_n=-\frac{1}{4}\alpha_n(\omega){\cal E }^2,
\end{eqnarray}
where, $\alpha_n(\omega)$ is known as the frequency dependent dipole polarizability, and is expressed as
\begin{eqnarray}
\label{eq2}
\alpha_n(\omega)&=& -[\langle \gamma _n J_n M_{J_n} | (\hat{\mbox{\boldmath$\varepsilon$}}^*\cdot{\bf D}) R_{n}^+(\omega) (\hat{\mbox{\boldmath$\varepsilon$}}\cdot{\bf D})\nonumber\\
&+&(\hat{\mbox{\boldmath$\varepsilon$}}\cdot{\bf D}) R_{n}^-(\omega) (\hat{\mbox{\boldmath$\varepsilon$}}^*\cdot{\bf D}) |\gamma _n J_n M_{J_n}\rangle],
\end{eqnarray}
where ${\bf D}=D \hat{\bf r}=- e \sum_j {\bf r}_j$ is the electric dipole (E1) operator with position of an $j^{th}$ electron ${\bf r}_j$ and 
the projection operators $R_{n}^{\pm}(\omega)$ are given by
\begin{eqnarray}
 R_{n}^{\pm}(\omega) = \sum_{k} \frac{|\gamma _k J_k M_{J_k} \rangle \langle \gamma _k J_k M_{J_k} |}{E_n - E_k \pm \omega }  .
\end{eqnarray}
 In the above expressions, $E_n$ and $J_n$ is the energy and angular momentum of the state $|\Psi_n\rangle$( denoted by $|\gamma _n J_n M_{J_n}\rangle$ in above expression), respectively, and 
sum over $k$ represents all possible allowed intermediate states $|\Psi_k \rangle$( denoted by $|\gamma _k J_k M_{J_k}\rangle$ in above expression) with $E_k$ and $J_k$ being the corresponding energies and angular momenta. $M_{J_n}$ and $M_{J_k}$ are the magnetic components of corresponding angular momenta. $\gamma _n$ and $\gamma _k$ include all the remaining quantum numbers of the corresponding state. Since {\bf D} is a vector 
operator, we obtain three terms resulting from the scalar, vector and tensor components, respectively. Thus, it 
can be given as 
\begin{eqnarray}
\label{eq3}
\alpha_n(\omega)= C_0 \alpha_n^{(0)}(\omega)+ C_1 \alpha_n^{(1)}(\omega)+ C_2 \alpha_n^{(2)}(\omega),
\end{eqnarray}
where $\alpha_n^{(0)}$, $\alpha_n^{(1)}$ and $\alpha_n^{(2)}$ are known as scalar, vector and tensor polarizabilities. In a sum-over-states
approach, it yields
\begin{equation}
\alpha_n^{(0)}(\omega)= \sum_{k \ne n} W_{n}^{(0)}  \left [\frac{ |\langle \gamma _nJ_n||{\bf D}||\gamma _k J_k \rangle|^2}{E_n -E_k +\omega}+\frac{ |\langle \gamma _n J_n||{\bf D}||\gamma _k J_k \rangle|^2}{E_n-E_k-\omega}\right], \label{eqpolz1}
\end{equation}
\begin{equation}
 \alpha_n^{(1)}(\omega)= \sum_{k \ne n} W_{n,k}^{(1)}  \left [\frac{ |\langle \gamma _n J_n||{\bf D}||\gamma _k J_k \rangle|^2}{E_n -E_k +\omega}-\frac{ |\langle\gamma _n J_n||{\bf D}||\gamma _k J_k \rangle|^2}{E_n-E_k-\omega}\right], \label{eqpolz2}
\end{equation}
and
\begin{equation}
\alpha_n^{(2)}(\omega)= \sum_{k \ne n} W_{n,k}^{(2)}  \left [\frac{ |\langle \gamma _n J_n||{\bf D}||\gamma _k J_k \rangle|^2}{E_n -E_k +\omega}+\frac{ |\langle \gamma _n J_n||{\bf D}||\gamma _k J_k \rangle|^2}{E_n-E_k-\omega}\right] \label{eqpolz3}
\end{equation}
with the coefficients
\begin{eqnarray}
W_{n}^{(0)} &=&-\frac{1}{3(2J_n+1)}, \label{eqp0}  \\
W_{n,k}^{(1)}&=&-\sqrt{\frac{6J_n}{(J_n+1)(2J_n+1)}}   \\
              & & \times (-1)^{J_n+J_k+1}  \left\{ \begin{array}{ccc}
                             J_n& 1 & J_n\\
                          1 & J_k &1 
                         \end{array}\right\},  \nonumber \label{eqp1}
\end{eqnarray}
and
\begin{eqnarray}
W_{n,k}^{(2)} &=&2\sqrt{\frac{5J_n(2J_n-1)}{6(J_n+1)(2J_n+3)(2J_n+1)}} \nonumber \\ 
& & \times (-1)^{J_n+J_k+1}
                                  \left\{ \begin{array}{ccc}
                                            J_n& 2 & J_n\\
                                            1 & J_k &1 
                                           \end{array}\right\} ,          \label{eqp2}
\end{eqnarray} 
for $\langle \gamma _n J_n||{\bf D}||\gamma _k J_k \rangle$ being the electric dipole (E1) reduced
matrix elements. The values of $C_0$, $C_1$ and $C_2$ coefficients in the above 
expressions depends on polarization of the electric field. In a number of applications oscillating electric fields produced 
by lasers with different choice of polarization are used depending on the suitability of an experimental geometry, but  most commonly
linearly and circularly polarized electric fields are considered. For linearly polarized light, one gets \cite{manakov}
\begin{eqnarray}
\label{eqlinear}
C_0 =1, \ \ \ C_1 =0, \ \ \ \text{and} \ \ \ C_2 = \frac{3M_J^2-J_n(J_n+1)}{J_n(2J_n-1)}
\end{eqnarray}
for the magnetic component $M_J$ of $J_n$. Here, it is assumed that the quantization axis is along the direction of polarization vector. Similarly, for circularly polarized light it corresponds to
\begin{eqnarray}
\label{eq}
C_0 =1, \ \  C_1=\frac{AM_J}{2J_n}, \ \  \text{and} \ \ C_2=-\frac{3M_J^2-J_n(J_n+1)}{2J_n(2J_n-1)},
\end{eqnarray}
where $A$ is known as the degree of circular polarization which possess the values $1$ and $-1$ for the right handed and left handed 
circularly polarized electric field, respectively and it is assumed in this case that the direction of quantization is along the wave vector.

The differential Stark shift of a transition $|\Psi_i \rangle \rightarrow |\Psi_f \rangle$ between an initial state $|\Psi_i\rangle$ to 
a final state $|\Psi_f \rangle$ is the difference between the Stark shifts of these states and is given by
\begin{eqnarray}\nonumber
\delta (\Delta E)_{if} (\omega ) &=&\Delta E_i(\omega)-\Delta E_f (\omega) \\
                      &=&-\frac{1}{4}\left[\alpha_i(\omega)-\alpha_f(\omega)\right]{\cal E}^2 .
\end{eqnarray} 
Differential stark shift between two states for linearly polarized light can be expressed as
\begin{eqnarray}\nonumber
\delta (\Delta E)_{if} (\omega )&=&-\frac{1}{4}\big [ \big\{ \alpha_i^{(0)}(\omega)-\alpha_f^{(0)}(\omega)\big \}+\\ \nonumber
                        & &\big \{\frac{3M_{J_i}^2-J_i(J_i+1)}{J_i(2J_i-1)} \alpha_i^{(2)}(\omega)-\\  
                        & & \frac{3M_{J_f}^2-J_f(J_f+1)}{J_f(2J_f-1)} \alpha_f^{(2)}(\omega) \big \}\big ]{\cal E}^2
\end{eqnarray} 
Similarly, differential stark shift between two states ($|\Psi_i \rangle$ and $|\Psi_f \rangle$) using circularly polarized light can be written as  
\begin{eqnarray}\nonumber
\delta (\Delta E)_{if} (\omega )&=&-\frac{1}{4}\big [ \big\{ \alpha_i^{(0)}(\omega)-\alpha_f^{(0)}(\omega)\big \}+ \\ \nonumber                           		    & &A\big\{\frac{M_{J_i}}{2J_i}\alpha_i^{(1)}(\omega)-\frac{M_{J_f}}{2J_f}\alpha_f^{(1)}(\omega)\big\}-\\ \nonumber
                        & &\frac{1}{2}\big \{\frac{3M_{J_i}^2-J_i(J_i+1)}{J_i(2J_i-1)} \alpha_i^{(2)}(\omega)-\\  
                        & & \frac{3M_{J_f}^2-J_f(J_f+1)}{J_f(2J_f-1)} \alpha_f^{(2)}(\omega) \big \}\big ]{\cal E}^2
\end{eqnarray} 
Here, $A=1$ when right circularly polarized light is used and $A=-1$ when left circularly polarized light is used. 
For a $\omega$ value at which $\delta (\Delta E)_{if} (\omega )$ is zero, the corresponding wavelength is a $\rm\lambda_{magic}$. 
Equivalently, it means finding out where the condition $\alpha_i(\omega)=\alpha_f(\omega)$ is satisfied for amplitude $\mbox{\boldmath${\cal E}$}$.

\begin{figure}
\centering 
\includegraphics[width=\columnwidth,keepaspectratio]{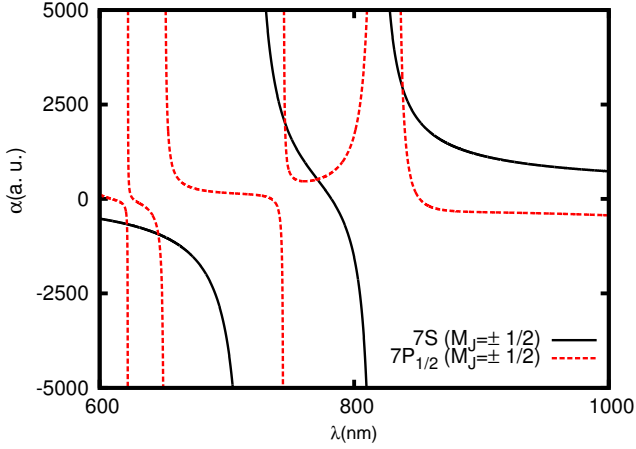}
\caption{(Color online) Dynamic polarizabilities (in a.u.) for the $7S_{1/2}$ and $7P_{1/2}$ states of Fr in the wavelength range 600-1000 nm for linearly polarized light
.}
\label{magiclinear1}
\end{figure} 

\begin{figure}
\centering 
\includegraphics[width=\columnwidth,keepaspectratio]{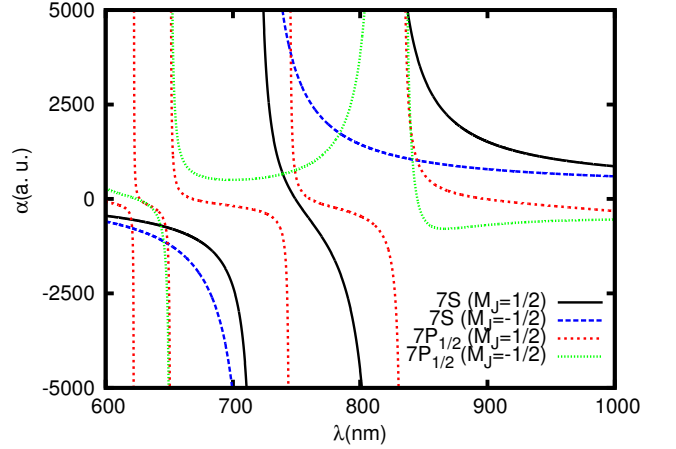}
\caption{(Color online) Dynamic polarizabilities (in a.u.) for the $7S_{1/2}$ and $7P_{1/2}$ states of Fr in the wavelength range 600-1000 nm for left circularly polarized light(A=-1).}
\label{magiccirc1}
\end{figure} 

\section{Method of Evaluation for Polarizability}  

To evaluate atomic wave functions of the ground and many low-lying excited states having a common closed core configuration $[6p^6]$
and a valence orbital (say, $n$) in Fr in the RCC theory framework, we first calculate the Dirac-Fock (DF) wave function ($|\Phi_0\rangle$)
for the closed core and then define a new working DF wave function of the entire state artificially as $\vert \Phi_n \rangle
= a_n^{\dagger}|\Phi_0\rangle$, appending the corresponding valence orbital $n$. In this procedure, evaluation of the exact atomic 
wave functions of the respective states requires incorporating the correlations among the electrons within $|\Phi_0\rangle$ that is 
referred to as core correlation, correlations effectively seen by only the valence electron of $|\Phi_n\rangle$ which is termed as 
valence correlation, and the correlations between the core electrons with the valence electron $v$ giving rise to core-valence correlation 
contributions. Using the wave operator formalism, one can express these wave functions accounting the above mentioned correlations 
individually as \cite{nandy,jasmeet}
\begin{eqnarray}
|\Psi_n \rangle &=& a_n^{\dagger} \Omega_c |\Phi_0 \rangle + \Omega_{cv} |\Phi_n \rangle + \Omega_v |\Phi_n \rangle,
\label{eqn21}
\end{eqnarray}
where $\Omega_c$, $\Omega_{cv}$ and $\Omega_v$ are known as the wave operators for the core ($c$), core-valence ($cv$) and valence 
($v$) correlations, respectively. As given in Eqs. (\ref{eqpolz1}) and (\ref{eqpolz2}), evaluation of $\alpha_n^{(i=0,1,2)}$
requires calculations of $|\langle \gamma _n J_n || {\bf D} ||\gamma _k J_k\rangle|^2$. Following the above conviction to classify correlation contributions,
we can express (see appendix of Ref. \cite{nandy})
\begin{eqnarray}
 \sum_k |\langle \gamma _n J_n || {\bf D} ||\gamma _k J_k\rangle |^2 = D^2_c +D^2_{cv} + D^2_v ,
\end{eqnarray}
 where $D^2_c$, $D^2_{cv}$ and $D^2_v$ are the contributions from the respective core, core-valence and valence 
correlations, respectively. Therefore, we can write
\begin{eqnarray}
\alpha_n^{(i)}  &=& \alpha_{n,c}^{(i)} + \alpha_{n,cv}^{(i)} + \alpha_{n,v}^{(i)}
\label{eq26}
\end{eqnarray}
for each component $i=0,1,2$ of $\alpha_n^{(i)}$. 

It can be later followed that $\alpha_{n,v}^{(i)}$ contribute the most in the evaluation of $\alpha_n$ in the considered states of Fr.
This contribution  can be effortlessly estimated to very high accuracy in the sum-over-states approach using the formula 
\begin{eqnarray}
\alpha_{n,v}^{(0)} (\omega) &=& 2 \sum_{k>N_c, k \ne n }^I W^{(0)}_{n}\frac{ (E_n-E_k) |\langle \gamma _n J_n || {\bf D}  || \gamma _k J_k\rangle |^2} 
{(E_n - E_k)^2 - \omega^2}, \nonumber \\
\end{eqnarray}

\begin{eqnarray}
\alpha_{n,v}^{(1)} (\omega) &=& -2\omega \sum_{k> N_c, k \ne n}^I W^{(1)}_{n,k}\frac{  |\langle \gamma _n J_n || {\bf D}  || \gamma _k J_k\rangle |^2} 
{(E_n - E_k)^2 - \omega^2}, \ \ \
\end{eqnarray}
and 
\begin{eqnarray}
\alpha_{n,v}^{(2)} (\omega) &=& 2 \sum_{k>N_c, k \ne n }^I W^{(2)}_{n,k}\frac{ (E_n-E_k) |\langle \gamma _n J_n || {\bf D}  || \gamma _k J_k\rangle |^2} 
{(E_n - E_k)^2 - \omega^2}, \nonumber \\
\end{eqnarray}
by calculating E1 matrix elements between the state of interest $|\Psi_n\rangle$ and many singly excited states $|\Psi_k\rangle$s
having common closed core with $|\Psi_n\rangle$. In the above equations, sum is restricted by involving states denoted by $k$ after 
$N_c$ and up to $I$, where $N_c$ represents for the core orbitals and $I$ represents for the bound states up to which we can determine
the $\langle \gamma _n J_n || {\bf D}  || \gamma _k J_k\rangle$ matrix elements explicitly in our calculation. 

In the RCC {\it ansatz}, these states can be commonly expressed for $|\Psi_n\rangle$ as \cite{Mukherjee10,bijaya1,bijaya2,ref1,ref2,bijaya3}
\begin{eqnarray}{\label{wav}}
|\Psi_n\rangle &=& e^T\{1+S_n\}|\Phi_n\rangle , \nonumber 
\end{eqnarray}
where the operators $T$ and $S_n$ are responsible for accounting core and valence correlations by exciting electrons from the core 
orbitals and valence orbital along with from the core orbitals, respectively. It can be noted that the core-valence correlations are 
accounted together by the simultaneous operations of $a_n^{\dagger}$ and $T$ as well as $S_n$ and $T$ operators. Since amplitudes 
of the $T$ and $S_n$ RCC operators are solved using coupled equations, the core and valence correlation effects together finally 
revamp quality of the wave functions.

In our calculations we have considered, all possible singly and doubly excited configurations (CCSD method) in the calculations of the amplitudes of the wave operators
$T$ and $S_n$. We have also included important triply excited configurations involving valence electron to elevate amplitudes of the 
RCC operators in the CCSD method wave operators (known as CCSD(T) method) in a perturbative approach as discussed in Ref. \cite{ref1}.

After obtaining the wave functions in the CCSD(T) method, we calculate E1 matrix element for a transition between the 
states $|\Psi_n \rangle$ and $|\Psi_k\rangle$ by evaluating the expression
\begin{eqnarray}
\langle \Psi_n| D| \Psi_k\rangle &=& \frac{\langle\Phi_n|\tilde{D}_{nk}|\Phi_k\rangle}{\sqrt{\langle\Phi_n|\{1+\tilde{N}_n\}|\Phi_n\rangle 
\langle\Phi_k|\{1+\tilde{N}_k\}|\Phi_k\rangle}} , \nonumber \\
\end{eqnarray}
 where $\tilde{D}_{nk}=\{1+S_n^{\dagger} \} e^{T^{\dagger}} D e^T \{1+S_{k}\}$ and $\tilde{N}_{i=n,k}=\{1+S_i^{\dagger} \} e^{T^{\dagger}}
e^T \{1+S_{i}\}$. 

In the above approach, it is only possible to take into account contributions only from the E1 matrix elements among the low-lying 
states to $\alpha_n^{(i)}(v)$ and refer to as ``Main($\alpha_{n,v}^{(i)}$)''. Contributions from higher excited states 
including continuum to $\alpha_n^{(i)}(v)$, denoted as ``Tail($\alpha_{n,v}^{(i)}$)'', are estimated approximately in the DF 
method using the expression
\begin{eqnarray}
\alpha_{n,v}^{(0)} ( \omega) &=& 2 \sum_{k > I} W^{(0)}_{n} \frac{ (\epsilon_n-\epsilon_k) |\langle \gamma _n
J_n || {\bf D} || \gamma _k J_k \rangle_{DF} |^2} {(\epsilon_n - \epsilon_k)^2 - \omega^2}, \ \ \ \ \ \
\end{eqnarray}

\begin{eqnarray}
\alpha_{n,v}^{(1)} ( \omega) &=& -2\omega \sum_{k > I} W^{(1)}_{n,k} \frac{ |\langle 
\gamma _n J_n || {\bf D} || \gamma _k J_k \rangle_{DF} |^2} {(\epsilon_n - \epsilon_k)^2 - \omega^2}, \ \ \ \ \ \
\end{eqnarray}
and 

\begin{eqnarray}
\alpha_{n,v}^{(2)} ( \omega) &=& 2 \sum_{k > I} W^{(2)}_{n,k} \frac{ (\epsilon_n-\epsilon_k) |\langle \gamma _n
J_n || {\bf D} || \gamma _k J_k \rangle_{DF} |^2} {(\epsilon_n - \epsilon_k)^2 - \omega^2}, \ \ \ \ \ \
\end{eqnarray}
where $\langle \gamma _n J_n || {\bf D} || \gamma _k J_k \rangle_{DF}$ are obtained using the DF wave functions, $k>I$ corresponds to the excited states including 
continuum whose matrix elements are not accounted in the Main($\alpha_{n,v}^{(i)}$) contribution, and $\epsilon$s are the DF energies. 
 
Similarly, the core-valence contributions $\alpha_{n,cv}^{(0)}$ is obtained at the DF method approximation using the expression
\begin{eqnarray}
\alpha_{n,cv}^{(0)} ( \omega) &=& 2 \sum_{k}^{N_c} W_{n}^{(0)} \frac{ (\epsilon_n-\epsilon_k) |\langle 
\gamma _n J_n || {\bf D} || \gamma _k J_k \rangle_{DF} |^2} {(\epsilon_n - \epsilon_k)^2 - \omega^2}, \nonumber \\
\end{eqnarray}
 
\begin{eqnarray}
\alpha_{n,cv}^{(1)} ( \omega) &=& -2\omega \sum_{k}^{N_c} W_{n,k}^{(1)} \frac{ |\langle 
\gamma _n J_n || {\bf D} ||\gamma _k J_k \rangle_{DF} |^2} {(\epsilon_n - \epsilon_k)^2 - \omega^2}. \ \ \ \ \ \
\end{eqnarray}
and
\begin{eqnarray}
\alpha_{n,cv}^{(2)} ( \omega) &=& 2 \sum_{k}^{N_c} W_{n,k}^{(2)} \frac{ (\epsilon_n-\epsilon_k) |\langle 
\gamma _n J_n || {\bf D} ||\gamma _k J_k \rangle_{DF} |^2} {(\epsilon_n - \epsilon_k)^2 - \omega^2}, \nonumber \\
\end{eqnarray}
We adopt a relativistic random phase approximation (RPA method), as discussed in Ref.~\cite{jasmeet,yashpal}, to evaluate 
$\alpha_{n,c}^{(0)}$ from the closed core of Fr. 

\begin{figure*}
\centering `
\includegraphics[scale=1.3]{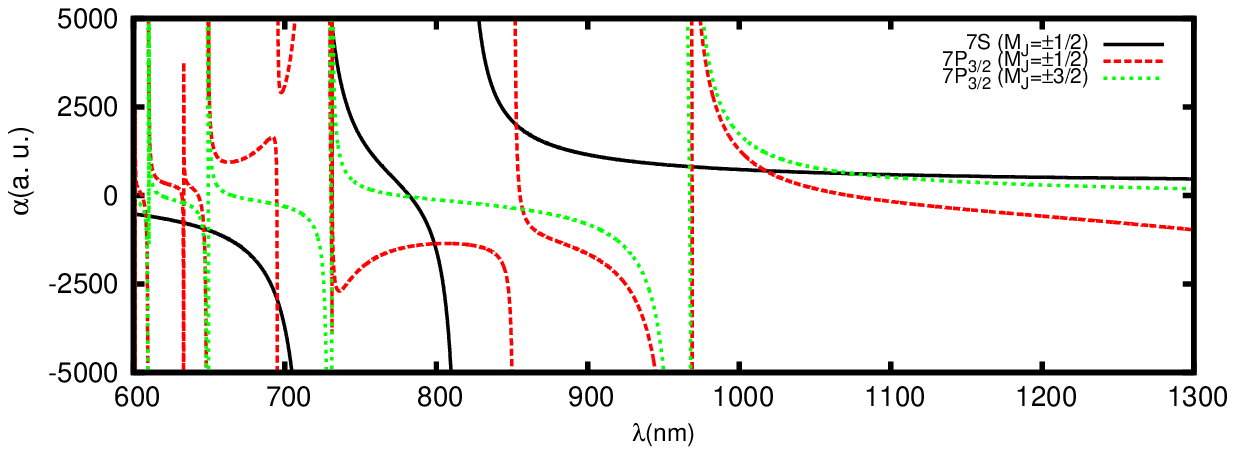}
\caption{(Color online) Dynamic polarizabilities (in a.u.) for the $7S_{1/2}$ and $7P_{3/2}$ states of Fr in the wavelength range 600-1300 nm for linearly polarized light.}
\label{magic3linear1}
\end{figure*}

\begin{figure*}
\centering 
\includegraphics[scale=1.3]{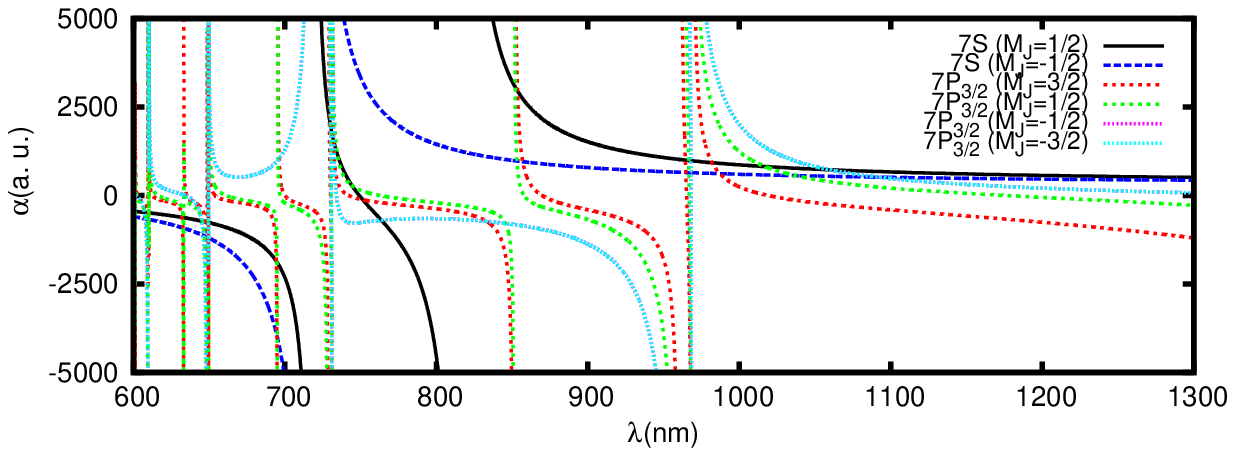}
\caption{(Color online) Dynamic polarizabilities (in a.u.) for the $7S_{1/2}$ and $7P_{3/2}$ states of Fr in the wavelength range 600-1300 nm for left circularly polarized light(A=-1).}
\label{magic3circ1}
\end{figure*}

\begin{table}
\caption{\label{magic1}Magic wavelengths ($\lambda _{\rm {magic}}$) (in nm) with corresponding polarizabilities ($\alpha_n(\omega)$) (in a.u.)
for the $7S-7P_{1/2}$ transition in the Fr atom with linearly polarized light along with the resonant wavelengths 
($\lambda_{\rm{res}}$) (in nm). Values that are found to be in discrepancy with $\lambda_{\rm{magic}}$ given in Ref. \cite{Dammalapati}
are highlighted in bold fonts.}
%\begin{ruledtabular}
\begin{tabular}{ccccc}
\hline  
 \hline
&&\multicolumn{2}{c}{Present}& Ref. \cite{Dammalapati}\\
&&\multicolumn{3}{c}{$M_j$=$\pm 1/2$} \\
Resonance &  $\lambda_{\rm{res}}$    &   $\lambda _{\rm {magic}}$   &   $\alpha_n(\omega)$ &  $\lambda _{\rm {magic}}$ \\
\hline
 & & & &  \\
&& 621.48 & -667 & 621.11 \\
$7P_{1/2}-10S_{1/2}$ &622.15 &&&\\ 
&& {\bf 646.05} & {\bf -930} & {\bf 642.85}  \\
$7P_{1/2}-8D_{3/2}$&650.9 &&&\\ 
$7s_{1/2}-7P_{3/2}$&718.18 &&&\\
$7P_{1/2}-9S_{1/2}$&744.4 &&&\\
&& 745.6    & 2015 & 745.36\\
&& {\bf 771.03}  & {\bf 520} & {\bf 797.75}\\
$7S_{1/2}-7P_{1/2}$&817.17 &&&\\
$7P_{1/2}-7D_{3/2}$&832.87 &&&\\
&& {\bf 838.08}  & {\bf 2933} & {\bf 871.62}  \\ 
$7P_{1/2}-8S_{1/2}$ &1332.87 &&&\\
&& {\bf 1479.49}  & {\bf 421} & \\       

\hline
\hline 
\end{tabular}
%\end{ruledtabular}
\end{table}

\begin{table*}
\caption{\label{magic12}Magic wavelengths ($\lambda _{\rm {magic}}$) (in nm) with corresponding polarizabilities ($\alpha_n(\omega)$) (in a.u.)
for the $7S-7P_{1/2}$ transition in the Fr atom with circularly polarized light along with the resonant wavelengths 
($\lambda_{\rm{res}}$) (in nm). Our values are compared with the corresponding $\lambda _{\rm {magic}}$ values given in Ref. \cite{Dammalapati}.}
%\begin{ruledtabular}
\begin{tabular}{ccccccccccc}
\hline  
 \hline
  & &  \multicolumn{5}{c}{Transition: $7S(M_{J}=1/2)-7P_{1/2}$}& \multicolumn{4}{c}{Transition: $7S(M_{J}=-1/2)-7P_{1/2}$}\\
   & &\multicolumn{3}{c}{$M_{J}=1/2$}&\multicolumn{2}{c}{$M_{J}=-1/2$}&\multicolumn{2}{c}{$M_{J}=1/2$}&\multicolumn{2}{c}{$M_{J}=-1/2$}\\
&&\multicolumn{2}{c}{Present}&Ref. \cite{Dammalapati}&&&&\\
Resonance &  $\lambda_{\rm{res}}$   &  $\lambda _{\rm {magic}}$   &   $\alpha_n(\omega)$ &  $\lambda _{\rm {magic}}$   & $\lambda _{\rm {magic}}$   &   $\alpha_n(\omega)$  &  $\lambda _{\rm {magic}}$   &   $\alpha_n(\omega)$ & $\lambda _{\rm {magic}}$   &   $\alpha_n(\omega)$ \\
\hline
 & & & & & & & \\
&&   620.10 & -543 &    &     &      &  620.84   &  -774   &   &    \\ 
$7P_{1/2}-10S_{1/2}$ & 622.15 &&&&&&&&&\\ 
&&   647.12 &  -741 &  640.25  & 643.24 &  -705  &  648.62 &  -1177 &  645.53  &  -1118   \\
$7P_{1/2}-8D_{3/2}$& 650.9 &&&&&&&&&\\ 
$7s_{1/2}-7P_{3/2}$& 718.18 &&&&&&&&&\\
&&        && &  739.24  &  662   &  & &&        \\
$7P_{1/2}-9S_{1/2}$&744.4 &&&&&&&&&\\
& &&&&&&&& 783.85& 1741\\
$7S_{1/2}-7P_{1/2}$& 817.17 &&&&&&&&&\\
$7P_{1/2}-7D_{3/2}$& 832.87 &&&&&&&&&\\
&&   835.31 & 5459  &  1116.2 &  837.08 &  5017  &  844.72 &  1022 & 841.10 &  1045  \\ 
$7P_{1/2}-8S_{1/2}$ &1332.87&&&&&&&&&\\       

\hline
\hline 
\end{tabular}
%\end{ruledtabular}
\end{table*}

\begin{table}
\caption{\label{magic31} Magic wavelengths ($\lambda _{\rm {magic}}$) (in nm) with corresponding polarizabilities ($\alpha_n(\omega)$) (in a.u.) 
for the $7S-7P_{3/2}$ transition in the Fr atom with linearly polarized light along with the resonant wavelengths 
($\lambda_{\rm{res}}$) (in nm) and their comparison with the $\lambda _{\rm {magic}}$ values given in Ref. \cite{Dammalapati}. Values showing large 
differences are shown in bold fonts.}
\scalebox{0.8}
%\begin{ruledtabular}
{\begin{tabular}{cccccccc}
\hline 
\hline
% & & & & & & & & & \\
%Transition&&\multicolumn{4}{c}{Linear Polarization}&\multicolumn{4}{c}{Circular Polarization}&\multicolumn{2}{c}{In presence of B}&Ref.\\ 
  \multicolumn{2}{c}{Transition}&&&&&\\
%\hline
\multicolumn{2}{c}{$7S(M_{J}=\pm 1/2)-7P_{3/2}$}&\multicolumn{3}{c}{$M_{J}=\pm 1/2$}&\multicolumn{3}{c}{$M_{J}=\pm 3/2$}\\
&&\multicolumn{2}{c}{Present}& Ref. \cite{Dammalapati}&\multicolumn{2}{c}{Present}&Ref. \cite{Dammalapati}\\
Resonance &  $\lambda_{\rm{res}}$   &   $\lambda _{\rm {magic}}$   &   $\alpha_n(\omega)$ & $\lambda _{\rm {magic}}$& $\lambda _{\rm {magic}}$   &   $\alpha_n(\omega)$ & $\lambda _{\rm {magic}}$ \\
\hline
&& 600.89  & -527 & 600.33    &   & &     \\ 
$7P_{3/2}-12S_{1/2}$ &601.12 & & & \\ 
&& {\bf 608.15}  & {\bf -570} & {\bf 605.64}    &  607.59 & -566 &606.66 \\ 
$7P_{3/2}-10D_{5/2}$&609.70 &&&\\ 
&& {\bf 610.27}  & {\bf -584}  &  &  610.18 &  -583 & 610.20\\
$7P_{3/2}-10D_{3/2}$&610.28 &&&\\
&& 632.83  & -771 & 632.38 & & & \\  
$7P_{3/2}-11S_{1/2}$&633.12 &&&\\
&& 646.49  & -936  & 645.11 & 645.64 & -924 & 645.95\\
$7P_{3/2}-9D_{5/2}$ &648.60 &&&\\
&& 649.65  & -983 & 649.65 &649.50 & -981 & 649.51 \\  
$7P_{3/2}-9D_{3/2}$ &649.67 &&&\\
&& 694.92  & -2943 & 694.67  &  &     &  \\
$7P_{3/2}-10S_{1/2}$ &695.09 &&&\\          
$7S_{1/2}-7P_{3/2}$ &718.18 &&&\\
$7P_{3/2}-8D_{5/2}$ &728.79 &&&\\
&& 729.63 & 5469  & 730.51 &729.77 &  5400 &729.73  \\
$7P_{3/2}-8D_{3/2}$ &731.17 &&&\\
&& 731.21  & 4766  & 731.32 & 731.88 & 4512 & 731.77   \\
&& {\bf 798.74}  &{\bf -1363}  & {\bf 784.62} &  782.83 & -39 &  783.35   \\  
$7S_{1/2}-7P_{1/2}$ &817.17 &&&\\       
$7P_{3/2}-9S_{1/2}$ &851.28 &&&\\ 
&&  852.84  &  1990 & 853.93 & & &  \\ 
$7P_{3/2}-7D_{5/2}$ &960.71 &&&\\
&& 968.79  & 810  & 968.83  &967.03 & 816 & 967.19  \\  
$7P_{3/2}-7D_{3/2}$ &968.99 &&&\\
&& {\bf 1017.45}  & {\bf 698}  & {\bf 1266.3} & {\bf 1076.60} & {\bf 613} &{\bf 1117.7}    \\         
\hline
\hline 
\end{tabular}
%\end{ruledtabular}
}
\end{table}

\begin{table*}
\caption{\label{magic32} Magic wavelengths ($\lambda _{\rm {magic}}$) (in nm) with corresponding polarizabilities ($\alpha_n(\omega)$) (in a.u.) for 
the $7S(M_J=1/2)-7P_{3/2}$ transition in the Fr atom with left handed circularly polarized light ($A=-1$) along with the resonant 
wavelengths ($\lambda_{\rm{res}}$) (in nm). Our values are compared with the corresponding $\lambda _{\rm {magic}}$ values given in Ref. \cite{Dammalapati}.}
%\begin{ruledtabular}
\begin{center}
\begin{tabular}{ccccccccccccccccc}
\hline 
\hline
%Transition&&\multicolumn{4}{c}{Linear Polarization}&\multicolumn{4}{c}{Circular Polarization}&\multicolumn{2}{c}{In presence of B}&Ref.\\ 
 \multicolumn{2}{c}{Transition: $7S(M_J=1/2)-7P_{3/2}$}&\multicolumn{3}{c}{$M_J=3/2$}&\multicolumn{3}{c}{$M_J=1/2$}&\multicolumn{2}{c}{$M_J=-1/2$}&\multicolumn{2}{c}{$M_J=-3/2$}\\
 &&\multicolumn{2}{c}{Present}&Ref. \cite{Dammalapati}&\multicolumn{2}{c}{Present}&Ref. \cite{Dammalapati}&&&&&\\
 \hline
Resonance &  $\lambda_{\rm{res}}$   &   $\lambda _{\rm {magic}}$   &   $\alpha_n(\omega)$ & $\lambda _{\rm {magic}}$& $\lambda _{\rm {magic}}$   &   $\alpha_n(\omega)$ & $\lambda _{\rm {magic}}$ &$\lambda _{\rm {magic}}$   &   $\alpha_n(\omega)$ &  $\lambda _{\rm {magic}}$   &   $\alpha_n(\omega)$  \\
\hline
&& 600.34 & -446 &     & 600.87  & -449 &  &  &   &  &   \\ 
$7P_{3/2}-12S_{1/2}$ &601.12 &&&\\ 
&& 608.86  & -485 & 603.36    &   607.66 & -479  & 605.99 &  607.32  &  -477  &  607.48  &  -478  \\ 
$7P_{3/2}-10D_{5/2}$&609.70 &&&\\ 
&& 610.10  & -491  &          &  610.16 & -491  &  & 610.23 &  -491  &   & \\
$7P_{3/2}-10D_{3/2}$&610.28 &&&\\
&& 632.08  & -619 & & 632.76    & -623 &   &   &   &    \\  
$7P_{3/2}-11S_{1/2}$&633.12 &&&\\
&& 647.41  & -744  & 643.62 & 645.52 & -726  &  645.60 &645.11  &  -723   &  645.45   &  -726\\
$7P_{3/2}-9D_{5/2}$ &648.60 &&&\\
&& 649.38  & -763 & &  649.47 & -764 & 649.57& 649.58  &  -765 &  &    \\  
$7P_{3/2}-9D_{3/2}$ &649.67 &&&\\
&& 694.26  & -1851   && 694.82 & -1887 & &   &   &   &       \\
$7P_{3/2}-10S_{1/2}$ &695.09 &&&\\          
$7S_{1/2}-7P_{3/2}$ &718.18 &&&\\
$7P_{3/2}-8D_{5/2}$ &728.79 &&&\\
&&  729.34 & 2175  & 730.64 & 730.15 &  1960  &729.76 &  730.36  &  1914 &  730.36  &  1914   \\
$7P_{3/2}-8D_{3/2}$ &731.17 &&&\\
&& 732.10  & 1557  && 734.63 & 1163  & 731.43&  732.16  &  1547   &   &    \\  
&& 751.47  & -124  && 744.27 & 282  &&  763.69  &  -699   & 783.25  &  -1879  \\ 
$7S_{1/2}-7P_{1/2}$ &817.17 &&&\\       
$7P_{3/2}-9S_{1/2}$ &851.28 &&&\\ 
&&  853.60  &  2950 && 852.03 & 3063 & &   &   &   &     \\ 
$7P_{3/2}-7D_{5/2}$ &960.71 &&&\\
&& 964.41  & 999  && 966.63 & 989  &&  967.97  &  983   &   &    \\  
$7P_{3/2}-7D_{3/2}$ &968.99 &&&\\
&& 982.05  & 925  & &1017.02 & 817 & 1395.3 &   1059.66   &  726   &   1062.67   &  721   \\
\hline
\hline 
\end{tabular}
\end{center}
%\end{ruledtabular}
\end{table*}

\begin{table*}
\caption{\label{magic33} Magic wavelengths ($\lambda _{\rm {magic}}$) (in nm) with corresponding polarizabilities ($\alpha_n(\omega)$) (in a.u.) for 
the $7S(M_J=-1/2)-7P_{3/2}$ transition in the Fr atom with left handed circularly polarized light ($A=-1$) along with the resonant wavelengths 
($\lambda_{\rm{res}}$) (in nm).}
%\begin{ruledtabular}
\begin{center}
\begin{tabular}{ccccccccccccccccc}
\hline 
\hline
%Transition&&\multicolumn{4}{c}{Linear Polarization}&\multicolumn{4}{c}{Circular Polarization}&\multicolumn{2}{c}{In presence of B}&Ref.\\ 
 \multicolumn{2}{c}{Transition: $7S(M_J=-1/2)-7P_{3/2}$}&\multicolumn{2}{c}{$M_J=3/2$}&\multicolumn{2}{c}{$M_J=1/2$}&\multicolumn{2}{c}{$M_J=-1/2$}&\multicolumn{2}{c}{$M_J=-3/2$}\\
 \hline
Resonance &  $\lambda_{\rm{res}}$   &   $\lambda _{\rm {magic}}$   &   $\alpha_n(\omega)$ &  $\lambda _{\rm {magic}}$   &   $\alpha_n(\omega)$ & $\lambda _{\rm {magic}}$   &   $\alpha_n(\omega)$ &  $\lambda _{\rm {magic}}$   &   $\alpha_n(\omega)$  \\
\hline
&& 600.59  & -602     & 600.95  & -605 &  &  &   &     \\ 
$7P_{3/2}-12S_{1/2}$ &601.12 &&&\\ 
&& 609.16  & -668     &  608.30 & -661  &  607.86  &  -657  &  607.82  &  -657  \\ 
$7P_{3/2}-10D_{5/2}$&609.70 &&&\\ 
&& 610.13  & -676  & 610.17 & -676  & 610.23 &  -677  &   & \\
$7P_{3/2}-10D_{3/2}$&610.28 &&&\\
&& 632.50  & -913 & 632.91 & -918 &   &   &   &    \\  
$7P_{3/2}-11S_{1/2}$&633.12 &&&\\
&& 647.96  & -1165  & 646.81 & -1143  &  646.17  &  -1131   &  646.08   &  -1129\\
$7P_{3/2}-9D_{5/2}$ &648.60 &&&\\
&& 649.45  & -1196 & 649.50 & -1197 &  649.59  &  -1199 &  &    \\  
$7P_{3/2}-9D_{3/2}$ &649.67 &&&\\
&& 694.73  & -3959   & 694.97 & -4002  &   &   &   &       \\
$7P_{3/2}-10S_{1/2}$ &695.09 &&&\\          
$7S_{1/2}-7P_{3/2}$ &718.18 &&&\\
$7P_{3/2}-8D_{5/2}$ &728.79 &&&\\
&& 728.95 & 9382  & 729.28 &  9113  &  729.57  &  8890 &  729.72  &  8777   \\
$7P_{3/2}-8D_{3/2}$ &731.17 &&&\\
&& 731.33  & 7727  & 731.43 & 7669  &   731.35  &  7718   &   &    \\  
$7S_{1/2}-7P_{1/2}$ &817.17 &&&\\       
$7P_{3/2}-9S_{1/2}$ &851.28 &&&\\ 
&&  857.06  &  954 & 853.19 & 974  &   &   &   &     \\ 
$7P_{3/2}-7D_{5/2}$ &960.71 &&&\\
&& 964.69  & 648  & 966.73 & 645  &  968.01  &  643   &   &    \\  
$7P_{3/2}-7D_{3/2}$ &968.99 &&&\\
&& 987.2  & 616  & 1037.18 & 560 &   1092.01   &  517   &   1083.72   &  522   \\
\hline
\hline 
\end{tabular}
\end{center}
%\end{ruledtabular}
\end{table*}

\section{Results and Discussion}

Accurate determination of $\alpha_n$ is very crucial in predicting  $\rm\lambda_{magic}$ precisely. In order to reduce the 
uncertainties in estimation of ``Main($\alpha_{n,v}^{(i)}$)'' contributions of the ground and the first two excited $7P_{1/2,3/2}$ states 
of the considered Fr atom, we use the experimentally driven precise values of E1 matrix elements for the $7S-7P_{1/2}$ and $7S-7P_{3/2}$ 
transitions, which are extracted from the lifetime measurements of the $7P_{1/2,3/2}$ states \cite{Simsarian2} and determine as many 
as E1 matrix elements of the transitions involving the low-lying states up to $11P$, $11S$ and $10D$ using the CCSD(T) method. Further,
we improve the results by using excitation energies from the measurements as listed in the National Institute of Science and Technology 
(NIST) database~\cite{NIST}. In order to demonstrate role of various contributions to $\alpha_n$, we give individual contributions  
from different E1 matrix elements to ``Main($\alpha_{n,v}^{(i)}$)'', ``Tail($\alpha_{n,v}^{(i)}$)'', $\alpha_{n,c}^{(i)}$ and 
$\alpha_{n,cv}^{(i)}$ (where $i=0,2$) explicitly along with the net results in the evaluation of static polarizabilities 
($\omega=0$) in atomic unit (a.u.) in Table \ref{pol1}. Moreover, we verify validity of these results by comparing with the previously 
reported other precise calculations in Refs. \cite{Derevianko10, Lim10, Wijngaarden} since the experimental data for these results 
is not available.

As seen in Table \ref{pol1}, our calculated $\alpha_n(0)$ value of 316.8 a.u. for the ground state of Fr atom is in agreement 
with the $\alpha_n(0)$ value of 317.8(2.4) a.u., which is calculated by Derevianko {\it et al.}  using a relativistic all order 
method \cite{Derevianko10}. Lim {\it et al.} had also calculated the static polarizability for the ground state as 315.2 a.u. using another RCC 
method in finite gradient approach considering Douglas-Kroll Hamiltonian \cite{Lim10}. Our result matches well with this value as well,
indicating validity of our calculation. The value of $\alpha_n(0)$ for the $7P_{1/2}$ state is estimated to be 1225 a.u., which 
agrees with the one given by Wijngaarden {\it et al.} as 1106 a.u. \cite{Wijngaarden}. Similarly, our calculated values for the 
static scalar and 
tensor polarizabilities of the $7P_{3/2}$ state in Fr are obtained as 2304 a.u. and $-467$ a.u., respectively. These results are 
again in reasonable agreement with the respective values given by Wijngaarden {\it et al.} as 2102 a.u. and $-402.7$ a.u. 
respectively \cite{Wijngaarden}. The reason for minor differences between our values and those obtained by Wijngaarden {\it et al.}   
could be because of the fact that the later calculations were carried out in a semi-empirical approach with the Coulomb approximation, 
while our calculations are more rigorous. Nevertheless, reasonable agreement between our calculations and the values reported by other 
theoretical calculations using a variety of many-body methods \cite{Derevianko10, Lim10, Wijngaarden} ascertain that our static 
values of $\alpha_n$ are reliable enough; in fact, we estimate about 1\% accuracy in our static polarizability values.
Correspondingly, we expect that the dynamic polarizabilities evaluated in our calculations are also accurate 
enough to determine $\lambda_{\rm{magic}}$ for the $7S_{1/2}-7P_{1/2,3/2}$ transitions in Fr.

Now to fathom about the accuracies in the estimated $\alpha_n$'s of the considered states in Fr by Dammalapati \textit{et al.} \cite{Dammalapati}, 
we consider the E1 matrix elements referred in their paper and use the experimental energies to reproduce the corresponding static 
polarizability values. As discussed in Sec. II(A) and II(B) of Ref. \cite{Dammalapati}, they take into account the E1 matrix 
elements of the $7S-nP$ transitions with $n = 7 - 20$ from Ref. \cite{Sansonetti} to evaluate dynamic $\alpha_n$ of the ground state. 
Similarly, they consider E1 matrix elements of the $7P-nS$ transitions for $n = 9 - 20$ and $7P-nD$ transitions for $n = 7 - 20$ states 
to evaluate $\alpha$ of the $7P_{1/2,3/2}$ states. Using the above referred data we were also able to reproduce plots 
given in Figs. 1-3 of Ref. \cite{Dammalapati}. The $\alpha_n(0)$ values obtained from these quoted matrix elements are given in 
Table \ref{pol1}. These values are 289.8 a.u., 57.23 a.u. and 98.57 a.u. for the scalar polarizabilities of the ground, 
$7P_{1/2}$ and $7P_{3/2}$ states respectively, while it is equal to 63.23 a.u. for the tensor polarizability of the $7P_{3/2}$ state. 
Compared to other calculations and our results, the reproduced ground state values differ slightly and mainly due to the extra 
core correlation effect taken into account in our calculation. In contrast, we find huge differences in the $\alpha_n(0)$ values of 
the $7P$ excited states. In accordance with the explicit contributions given in Table \ref{pol1}, we observe that this discrepancy is 
mainly due to omission of the E1 matrix element contributions from the $7P_{1/2}-6D_{3/2}$ and $7P_{3/2}-6D_{5/2}$ transitions, which 
alone contribute more than 60\% to the total polarizabilities of the $7P_{1/2}$ and $7P_{3/2}$ states. This obviously implies that 
$\alpha_n$ used by Dammalapati \textit{et al.} are not reliable enough for determining $\lambda_{\rm{magic}}$ precisely.

In pursuance of demonstrating $\rm \lambda_{magic}$ for the $7S-7P_{1/2,3/2}$ transitions in Fr,  we plot the dynamic $\alpha_n$ values 
of the $7S$, $7P_{1/2}$ and $7P_{3/2}$ states in Figs. \ref{magiclinear1}, \ref{magiccirc1}, \ref{magic3linear1} and \ref{magic3circ1} 
for both linearly and circularly polarized light separately. The wavelengths at which this intersection takes place are identified as 
$\lambda_{\rm{magic}}$ and are listed in Tables \ref{magic1}, \ref{magic12}, \ref{magic31}, \ref{magic32} and \ref{magic33}. As
discussed in Ref.~\cite{arora1} the occurrence of $\lambda_{\rm{magic}}$ can be predicted between the resonant wavelengths $\lambda_{\rm{res}}$ which has also been listed in these tables along with the corresponding resonant transition. $\rm \lambda_{magic}$ are tabulated in rows lying between two resonances to identify
the placements of $\lambda_{\rm{magic}}$ clearly between two $\lambda_{\rm{res}}$. Below we
discuss these results for the $7S-7P_{1/2}$ and $7S-7P_{3/2}$ transitions separately for both linearly and circularly polarized 
light and highlight the discrepancies in our results from the results presented in Ref. {\citep{Dammalapati}}.

\subsection{$\rm \lambda_{magic}$ for the $7S-7P_{1/2}$ transition}

A total of six $\rm \lambda_{magic}$ for the $7S-7P_{1/2}$ transition using linearly polarized light are listed in Table~\ref{magic1}
in the wavelength range 600-1500 nm. Major differences found between our results from the values presented in Ref.~\cite{Dammalapati} are 
marked in bold font. A $\rm \lambda_{magic}$ reported at 642.85 nm in Ref. \citep{Dammalapati} is instead of found to be 
at 646.05 nm. Our analysis suggests this discrepancy is mainly due to different E1 amplitude of the $7P_{1/2}-8D_{3/2}$ transition 
obtained by the CCSD(T) method in the present work as compared to the one obtained  using the HFR method in Ref.~\cite{Dammalapati}. In near infrared region (i.e. 700-1200 nm), two out of three $\rm \lambda_{magic}$ 
are identified at different wavelengths using our method as compared to $\rm \lambda_{magic}$ reported by Dammalapati \textit{et al.}. 
This disagreement is mainly due to inclusion of the E1 amplitude of the $7S-7P_{1/2}$ transition in the present calculation of 
$7P_{1/2}$ polarizability which play crucial role in this region. As a consequence, we find a $\rm \lambda_{magic}$ at 771.03 nm
supporting a red detuned trapping scheme, which is evident from the positive sign of the polarizability values at this wavelength as
shown in Fig.~\ref{magiclinear1} and quoted in Table~\ref{magic1}. Instead this was reported at 797.75 by Dammalapati {\it et al.} and 
was seen supporting a blue detuned trap in Fig. 3(a) of Ref.~\cite{Dammalapati}, since the corresponding light shift value had 
positive sign at this wavelength. Similarly, the $\rm \lambda_{magic}$ for the $7S-7P_{1/2}$ transition using circularly polarized 
light are tabulated in Table~\ref{magic12} and graphically presented in Fig.~\ref{magiccirc1}. In the present work, we determine $\rm 
\lambda_{magic}$s for left circularly polarization using $A=-1$ considering all possible positive and negative $M_J$ sublevels of the 
states participating in the transition. Note that $\rm \lambda_{magic}$ for the right circularly polarized light of a transition with 
a given $M_J$ are equal to left circularly polarized light with opposite sign of $M_J$. From Table \ref{magic12}, we find large 
differences between $\rm \lambda_{magic}$ reported in Ref.~\cite{Dammalapati} and those obtained by us. 

\subsection{$\rm \lambda_{magic}$ for the $7S-7P_{3/2}$ transition}

 The $\rm \lambda_{magic}$ for the $7S-7P_{3/2}$ transition are identified from the crossings of the dynamic polarizabilities of 
the $7S$ and $7P_{3/2}$ states as shown from their plotting in Figs. \ref{magic3linear1} and \ref{magic3circ1} for both linearly and 
circularly polarized light respectively. These values are presented separately in Table \ref{magic31} for the $7S-7P_{3/2}$ transition 
using linearly polarized light while they are given in Tables \ref{magic32} and \ref{magic33} for the $7S(M_J=1/2)-7P_{3/2}$ and 
$7S(M_J=-1/2)-7P_{3/2}$ transitions, respectively, using circularly polarized light. At least four discrepancies among 
$\rm \lambda_{magic}$ are found in comparison to the values reported in Ref.~\cite{Dammalapati} and are highlighted in bold fonts in 
the above tables. The first disagreement is in the $\rm \lambda_{magic}$ value reported in this work at 608.15 nm in the vicinity of 
the $7P_{3/2}-10D_{5/2}$ transition, but was identified at 605.64 nm in Ref.~\cite{Dammalapati}. The reason for this disagreement is 
primarily due to the difference in the E1 matrix element for the $7P_{3/2}-10D_{5/2}$ transition used in both the works, which 
contributes significantly around this wavelength. As shown in Table~\ref{pol1}, the E1 matrix element for the $7P_{3/2}-10D_{5/2}$ 
transition obtained by the CCSD(T) method is 1.27 a.u., whereas, the value used by Dammalapati \textit{et al.} was 1.55 a.u.. From Table 
\ref{magic31}, it is also evident that we are able to identify one $\rm \lambda_{magic}$ for the $7S-7P_{3/2}(M_{J}=\pm 1/2)$ 
transition at 610.27 nm, there was no corresponding value was found in Ref.~\cite{Dammalapati}. Moreover, $\rm \lambda_{magic}$ 
for the above transition reported at 784.62 nm by Dammalapati \textit{et al.} in Fig. 2 of Ref.~\cite{Dammalapati} is close to the tune-out 
wavelength (wavelength at which the ac polarizability of the ground state becomes zero). As seen in Table~\ref{magic31}, the value of 
ac polarizability at the corresponding $\rm \lambda_{magic}$ at 798.74 nm comes out to be a large negative value in this work. Hence, 
the trap at this $\rm \lambda_{magic}$ indicate to support a strong blue detuned trap as compared to a shallow blue detuned trap 
portrayed in Ref. ~\cite{Dammalapati}. Similarly, our calculated and their reported $\rm \lambda_{magic}$ after the resonant transition
$7P_{3/2}-7D_{3/2}$ (beyond 968.99 nm) are completely different. This can be attributed to the fact that the resonant transitions which 
appear after 968.99 nm (i.e. $7P_{3/2}-8S_{1/2}$, $7P_{3/2}-6D_{5/2}$ and $7P_{3/2}-6D_{3/2}$ transitions) have not been taken into 
account by Dammalapati \textit{et al.} in their calculation of the $7P_{3/2}$ state polarizabilities. Furthermore, we have listed 
$\rm \lambda_{magic}$ for the $7S(M_J=1/2)-7P_{3/2}$ and $7S(M_J=-1/2)-7P_{3/2}$ transitions using circularly polarized light in Tables 
\ref{magic32} and \ref{magic33}. In this case too, we find more number of $\rm \lambda_{magic}$ and the ones reported by Dammalapati 
\textit{et al.} do not agree with our values at most of the places.

\section{Conclusion}

In summary, we present a list of recommended magic wavelengths for the $7S-7P_{1/2,3/2}$ transitions of the Fr atom considering 
both linearly and circularly polarized light, which will be very useful to trap Fr atoms at these wavelengths for high precision 
experiments. We have calculated dynamic electric dipole polarizabilities of the ground and $7P_{1/2,3/2}$ states of Fr by combining 
matrix elements  calculated using the precisely measured lifetimes of the $7P_{1/2,3/2}$ states and performing calculations of higher excited states
using a relativistic coupled-cluster method. Reliability of these results are verified by comparing the static dipole polarizability 
values with the other available theoretical results. Since experimental results of these quantities are not available, our calculations 
will serve as bench mark values for the future measurements. The magic wavelengths for these transitions were investigated earlier using electric dipole matrix elements from literature, but omitting many dominant contributions 
such as core correlation contribution and some very important E1 transitions. We 
present the revised values of the magic wavelengths of the above D-lines for both linearly and circularly polarized light in the 
optical region taking into account all the omitted contributions. We even highlight the discrepancy in the prediction of different kind of trap to be used at some magic wavelengths in the present work and as
interpreted from the previous study. These magic wavelengths will be of immense interest to the experimentalists to carry out cold atom 
experiments and investigating many fundamental physics using Fr atoms.
\\
\section*{Acknowledgements}
S.S. acknowledges financial support from UGC-BSR scheme. B.K.S acknowledges use of Vikram-100 HPC Cluster at Physical Research
Laboratory, Ahmedabad. The work of B.A. is supported by CSIR Grant No. 03(1268)/13/EMR-II, India. 

%\bibliography{refs.bib}

\end{document}